\pdfoutput=1
\documentclass[aps,prd,preprint,superscriptaddress]{revtex4}
\usepackage{graphicx}  % needed for figures
\usepackage{dcolumn}   % needed for some tables
\usepackage{bm}        % for math
\usepackage{amssymb}   % for math

\begin{document}

%A\begin{flushright}
%CDF/PUB/TOP/CDFR/3985 \\
%AVersion 1.04\\
%A\today \\
%A\end{flushright}

%\pagestyle{myheadings}
%A\begin{tabbing}
%Aaaaaaaaaaa\=aaaaa\=aaaaaaaaaaaaaaaaaaaaa\=      \kill
%A\\
%A\end{tabbing}
  
%A\title{Visualization Drivers for Geant4}
\title{
{ Visualization Drivers for Geant4 \\
 Fermilab-TM-2329-CD, Oct 2005}
}

\author{Andy Beretvas}
%\date{Oct. 7, 2005}
%A\maketitle
\begin{abstract}
This document is on Geant4 visualization tools (drivers), evaluating pros and
cons of each option, including recommendations on which tools to support at
Fermilab for different applications{\cite{Daniel}}.

Four visualization drivers are evaluated. They are OpenGL, HepRep,
DAWN and VRML. They all have good features, OpenGL provides graphic
output with out an intermediate file! HepRep provides menus to assist
the user. DAWN provides high quality plots and even for large files
produces output quickly. VRML uses the smallest disk space for intermediate
files.

Large experiments at Fermilab will want to write their own display.
They should proceed to make this display graphics independent.
Medium experiment will probably want to use HepRep 
because of it's menu support. Smaller scale experiments will
want to use OpenGL in the spirit of having immediate response, good quality
output and keeping things simple.
\end{abstract}

\maketitle
%1
   %==============================================================	
\section{Introduction}
%===============================================================

        This report start with the sections Introduction,
	Documentation, Available Drivers, followed by a tour of Visualization.
	The tour shows plots produced by Geant4 using four different device
	drivers. The next section is on Building OpenGL
	a graphics device driver that produces immediate results. Followed by two
	further sections on OpenGL. We then treat graphic device drivers that
	require intermediate files. These are HepRep, DAWN and VRML.
	An important subsection is `Summary of intermediate files'.
	The report concludes with `Which driver to use?'.
	
	Very important to any experiment is a good way to visualize
	the results.To illustrate this obvious point I have selected
	an example from High Energy Physics (HEP).
	 I was lucky to be able to play a part in
	the CDF display{~\cite{CDF_Display}}. 
	By looking at the central tracking display
	and seeing that there were very few high P$_{t}$ tracks
	in minimum-bias events,
	it was obvious that a fast trigger processor was needed
	for the experiment. The central tracking chamber is a cylindrical
	drift chamber located in a 1.5-T magnetic field provided by a
	super-conducting solenoid coaxial with the beam.
	Thus  high P$_{t}$ tracks are almost straight lines. 
	Jim Freeman and Bill Foster proposed and built the first
	fast trigger processor for the experiment.

	Equally true is that any experiment today needs a simulation.
	The total cross section  for $p\overline{p}$
	interactions at $\sqrt{s}$ = 1800 GeV
	was measured by two groups CDF and E811\cite{total}. 
	One used a simulation(CDF) and one did not(E811). 
	The results differ by about 2.8$\sigma$. I along with some of my
	CDF collaborators wrote a report that indicated it was much better to
	have the simulation\cite{CDF_E811}.

\begin{figure}[htbp]
\includegraphics{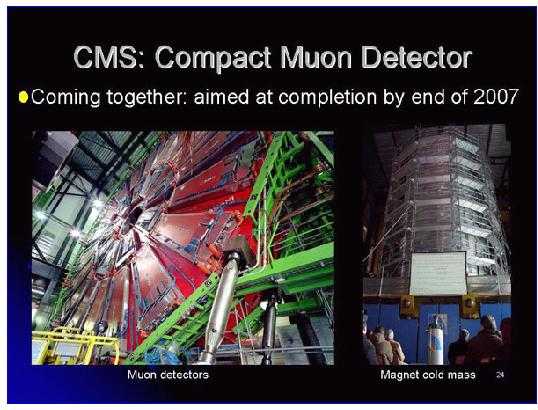}
\caption{\small
 This is slide \#24 from the talk by Pier Oddone
 showing the CMS detector.\newline
 The CMS collaboration has 2000 members of which 300 are in the US.
 }
\label{figure:CMS_detector}
\end{figure}

\begin{figure}[htbp]
\includegraphics[width=0.8\textwidth, clip]{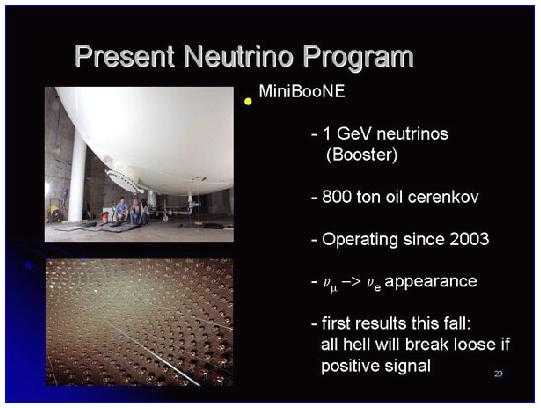}
\caption{\small
 This is slide \#29 from the talk by Pier Oddone
 showing the MiniBooNE detector.\newline
 The MiniBooNE detector has about 80 members.
 }
\label{figure:MiniBooNE_detector}
\end{figure}

	For any experiment it is always good if you can show
	a display and histogram for the experiment and the corresponding
	simulation. 
	The context of different applications is very broad at
	Fermilab. Clearly the most important item on our
	agenda is the Linear Collider. Pier Oddone explains this
	in his ``A vision for Fermilab''\cite{vision}. 
	 Physicist at SLAC are already
	using Geant4 in detector simulations for the Linear Collider.
	The CMS detector is shown in Fig.~\ref{figure:CMS_detector}.
The CMS experiment is currently rewriting it's software\cite{EDM}.
	The new software will use Geant4.
	The older software called OSCAR also uses Geant4\cite{OSCAR}.
        Geant4 has also been used to simulate
	a test of the Hadron Calorimeter which occurred in 2002.
	Of the current experiments MiniBooNE
	(see Fig.~\ref{figure:MiniBooNE_detector}) is using Geant4.

\begin{figure}[htbp]
\includegraphics[width=0.8\textwidth, clip]{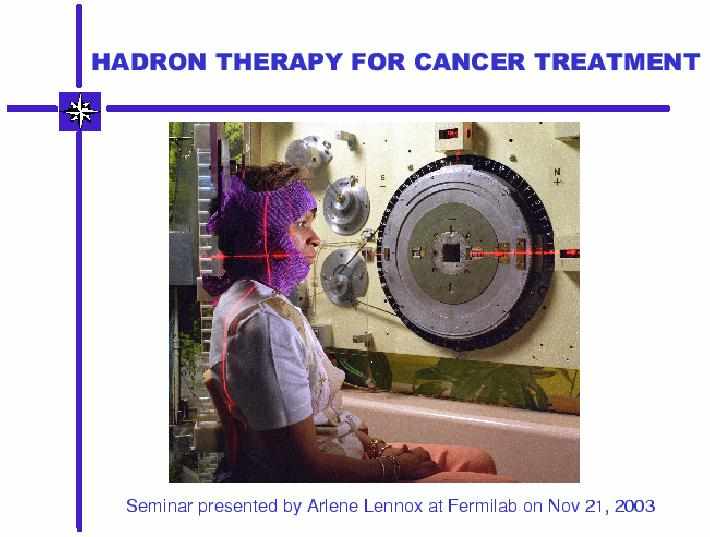}
\caption{\small
A very small group of people run the hadron therapy facility at Fermilab. }
\label{figure:hadron_therapy}
\end{figure}

	Some Fermilab physicist are involved in a planned experiment (MICE)
	at the Rutherford Appleton Laboratory  involving muon cooling.
	The beam line for MICE has been simulated using Geant4.
	Cancer patients have been treated at Fermilab using hadron
	therapy for many years\cite{Lennox}.
	Fig.\ref{figure:hadron_therapy} is the first slide 
	in a talk given by Arlene Lennox.

	Users at Fermilab include summer students working on medical
	applications\cite{Amanda}.
	To provide answers for visualization over such a broad range
	of applications is a difficult question.
	
	Geant4 has done a very good job of making this question
	a little easier to answer by factoring the problem into visualization
	commands and drivers.
	The basic idea is that the visualization commands do not depend
	on the driver. Further it was hoped that once the code was built
	one could use any driver.The design was for the graphics systems
	to be complementary to each other. 
	Because of the factorization this report will be about the different
	drivers.
%============================================================	
\section{Documentation}	
%============================================================
	
	This report is based on running on the latest version of Geant4
	version v4\_7\_1. Some of my experience has been with earlier
	versions v4\_6\_2\_p02, v4\_7\_0, and v4\_7\_0\_p01.
	During this time the Geant4 graphics system has been stable.
	Version v4\_7\_1 was the first time I built an official Fermilab OpenGL
	version.

	To learn about Visualization in Geant4 I have used chapter 8
	(Visualization)of the Geant User's Guide for Application
	Developers\cite{UsersGuide}.
        The visualization chapter also covers visualization 
	of detector geometry trees.
	These are easy to use (RegisterGraphicsSystem (new G4ASCIITree);
	 Idle$>$ /vis/drawTree ! ATree), but are not relevant to 
	 our discussion on drivers.
         Visualization is also covered in a talk by Joseph Perl
	 presented as part of a Geant4 tutorial\cite{Perl}.

\section{Available Drivers}
	 The list of drivers is given in Table 8.1 of the User's Guide.
	 The 11 possible choices are DAWNFILE, DAWN-Network,
	 HepRepFile, OpenGL-Xlib, OpenGL-Motif, OpenGL-Win32, 
	 OpenInventor-X, OpenInventor-Win32, RayTracer, VRMLFILE,
	 VRML-Network.
	 
	 At Fermilab  only Unix products are supported. At CERN
	 there is support for running Geant4 under windows. Thus two choices
	 are eliminated (OpenGL-Win32, OpenInventor-Win32).
	 I have chosen not to investigate the network drivers.
	 They are DAWN-Network, and VRML-Network.
	 The OpenInventor drivers were developed at Fermilab by
	 Jeff Kallenbach. At the present time no one is maintaining them.
	 This eliminates OpenInventor-X from the list.
	 Information about the RayTracer program can be found 
	 in the standard Geant4 source code\cite{RayTracer}.
The documentation for RayTracer says that only a limited set of commands
are supported. 
Further, the  README says "G4RayTracer can visualize absolutely all
       kinds of geometrical shapes which G4Navigator can deal with.
        Instead, it can NOT visualize hits nor trajectories generated
	by usual simulation" 
	RayTracer it would seem would be valuable for finding problems 
	in very complicated geometries. Because of it's inability to
	deal with trajectories I have eliminated it. I have chosen 
	to investigate OpenGL-Xlib, rather than OpenGL-Motif. In summary
	there are 4 general purpose drivers that will be investigated.

	The remaining four drivers are: OpenGL-Xlib, VRMLFILE, 
	HepRepFile and DAWNFILE.
	Thus I have run novice example3 with all of the four remaining drivers. 
	Incidentally the talk by Joseph Perl is about three of the drivers
	(OpenGL, HepRepFile, DAWNFILE). The BTeV experiment (which included
	Rob Kutschke, Lynn Garren, Paul Lebrun,  Julia Yarba, ...) 
	would have used VRMLFILE.

\section{Tour of Visualization}
\subsection{OpenGL}
\begin{figure}[htbp]
     \includegraphics[width=0.8\textwidth, clip]{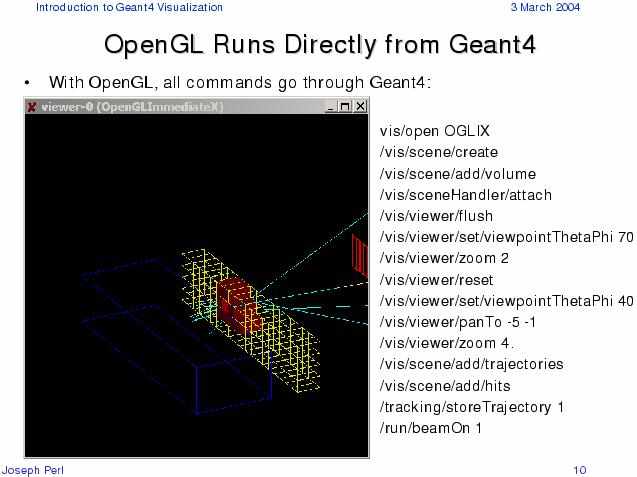}
\caption{\small
 This is an OpenGL plot shown by Joseph Perl at the SLAC Mar. 2004 tutorial.
 }
\label{figure:OpenGL_1}
\end{figure}

     We start our tour with an OpenGL plot shown in
     Fig.\ref{figure:OpenGL_1}.
 
     This is from a talk by Joseph Perl
     which was part of a Geant4 Tutorial held at SLAC in Mar.of 2004. 
     The slide shows some typical Geant4 visualization commands.
      The first command /vis/open
     OGLIX opens the visualization driver. The title is very important
     and indicates that OpenGL runs directly from Geant4. All other drivers
     (HepRepFile, DAWNFILE and VRMLFILE) produce an intermediate file.
     It's important to mention that the visualization code 
     can be C++ source code or visualization commands. It's very nice to be in
     the interactive mode and see the display change. Sometimes you may
     mistype a command and you learn that right away. It's also important to
     realize that if you have many commands you can put them in a macro file
     (vis.mac).
     These commands can be executed at once by giving the command
     /control/execute vis.mac. Thus you can proceed efficiently in the
     interactive mode.
     
\begin{figure}[htbp]
\includegraphics[width=0.8\textwidth, clip]{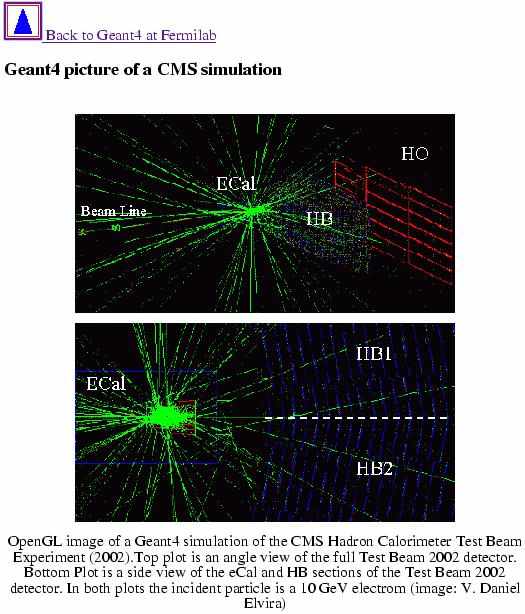}
\caption{\small
 This is an OpenGL plot shown by Daniel Elvira at the Fermilab Oct. 2003 tutorial.
 }
\label{figure:OpenGL_2}
\end{figure}

     OpenGL plots have also been produced at Fermilab. The following plot
     shown in Fig. \ref{figure:OpenGL_2}
     created by V. Daniel Elvira is taken from our new Geant4 web page\cite{New_Web_Page}. 

\begin{figure}[htbp]
\includegraphics[width=0.8\textwidth, clip]{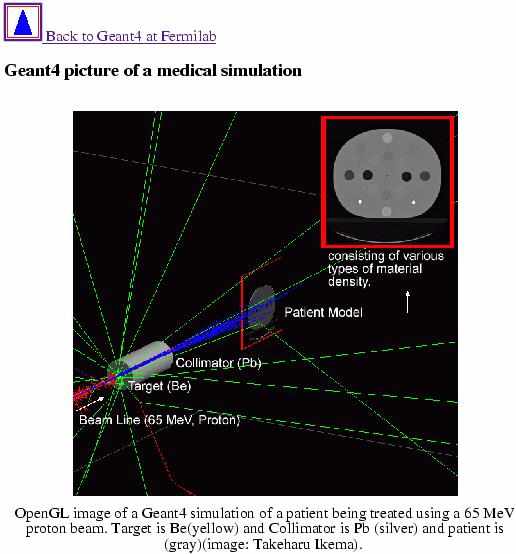}
\caption{\small
 This is an OpenGL plot produced by Takeharu Ikema.
 }
\label{figure:OpenGL_3}
\end{figure}

     The web page has a section called pictures.
     Daniel's plot is a simulation of an electron showering in the CMS
     EM calorimeter. Another OpenGL plot produced at Fermilab is shown
     in Fig.~\ref{figure:OpenGL_3}.

     This plot was produced by a visitor
     Takeharu Ikema who worked on Medical Physics Applications.

 \subsection{HepRep}    
\begin{figure}[htbp]
\includegraphics[width=0.8\textwidth, clip]{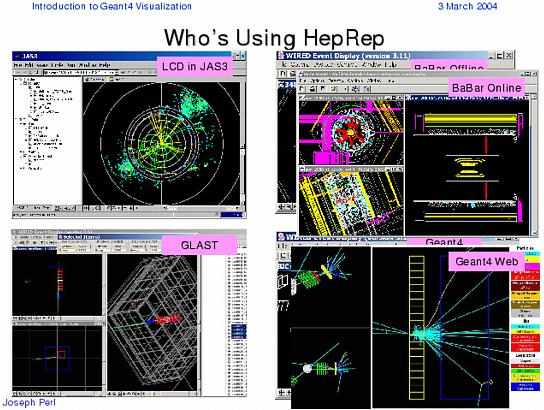}
\caption{\small
HepRep plots from a talk presented by Joseph Perl (Tutorial SLAC Mar. 2004).
}
\label{figure:HepRep_1}
\end{figure}

     We continue our tour by again looking at another plot 
     from Joseph Perl's talk (Fig.\ref{figure:HepRep_1}).

      This plot
     shows that HepRep is being used by many collaborations world wide.
      The plot in the upper left hand corner is made using JAS3 a
      general purpose, open-source data analysis tool. JAS is the abbreviation
      for JAVA Analysis Studio. The LCD indicates that this is a study 
      for the Linear Collider Detector. Another important application
      for HepRep is the BeBar b-experiment at SLAC shown in the upper
      right corner. HepRep also serves the space community. GLAST is
      the Gamma Ray Large Array Telescope.
      The GLAST simulation is shown in the lower left hand corner.
      GLAST is scheduled to be launched in 2006\cite{GLAST}.
      The lower right hand plot indicates that the Geant4 Web pages
      are maintained by using HepRep.
      
\begin{figure}[htbp]
\includegraphics[width=0.8\textwidth, clip]{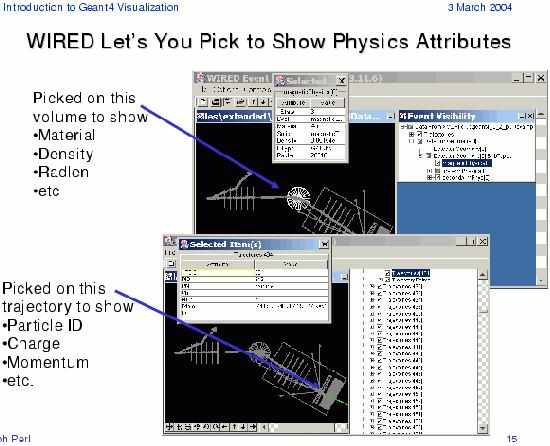}
\caption{\small
Picking important features of the event.\newline 
Also taken from Joseph Perl's talk at SLAC Mar. 2004. 
 }
\label{figure:Perl_15}
\end{figure}

      At Fermilab we (Andy Beretvas, V. Daniel Elvira
      and Panagiotis Spentzouris) have agreed to support HepRep.
      The reasons for our decision are that HepRep has menu's to guide
      the user and is easy to install. Fig. \ref{figure:Perl_15} explains 
      how using the WIRED
      client event display you can pick to show important features
      of the event (Particle ID, charge, momentum).

\begin{figure}[htbp]
\includegraphics[width=0.8\textwidth, clip]{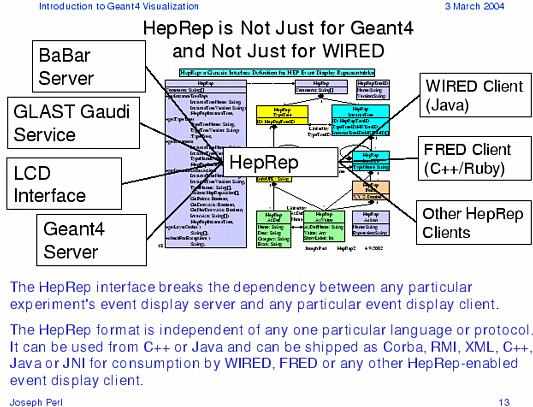}
\caption{\small
 HepRep is not just for Geant4 is the title of this slide.\newline
 Also taken from Joseph Perl's talk at SLAC Mar. 2004. 
 }
\label{figure:HepRep_all}
\end{figure}

      However, there are many reasons for this choice
      such as HepRep will work in a wider context. This is explained
      by Joseph Perl in  Fig.~\ref{figure:HepRep_all}.

 \subsection{DAWN}
\begin{figure}[htbp]
\includegraphics[width=0.8\textwidth, clip]{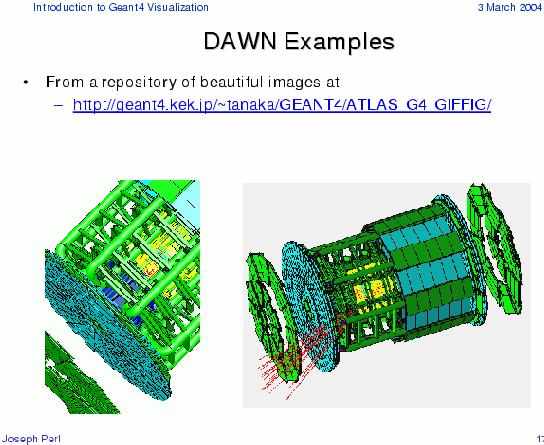}
\caption{\small
DAWN is used to display a simulation of the ATLAS detector. \newline
Also taken from Joseph Perl's talk at SLAC Mar. 2004. 
 }
\label{figure:DAWN_ATLAS}
\end{figure}

      The next plot Fig.~\ref{figure:DAWN_ATLAS} shows a
      simulation of the ATLAS detector. 
      Again the figure is taken from the
      talk by Joseph Perl. From the drawing the size of the detector is not
      clear. The detector is huge with a length of 45m and a height of 22m
      and contains more than million components\cite{Atlas_detector}.
     
      I do not know of any Geant4 users at Fermilab who use DAWNFILE as a
      visualization driver.
      
\subsection{VRML} 
\begin{figure}[htbp]
\includegraphics[width=0.8\textwidth, clip]{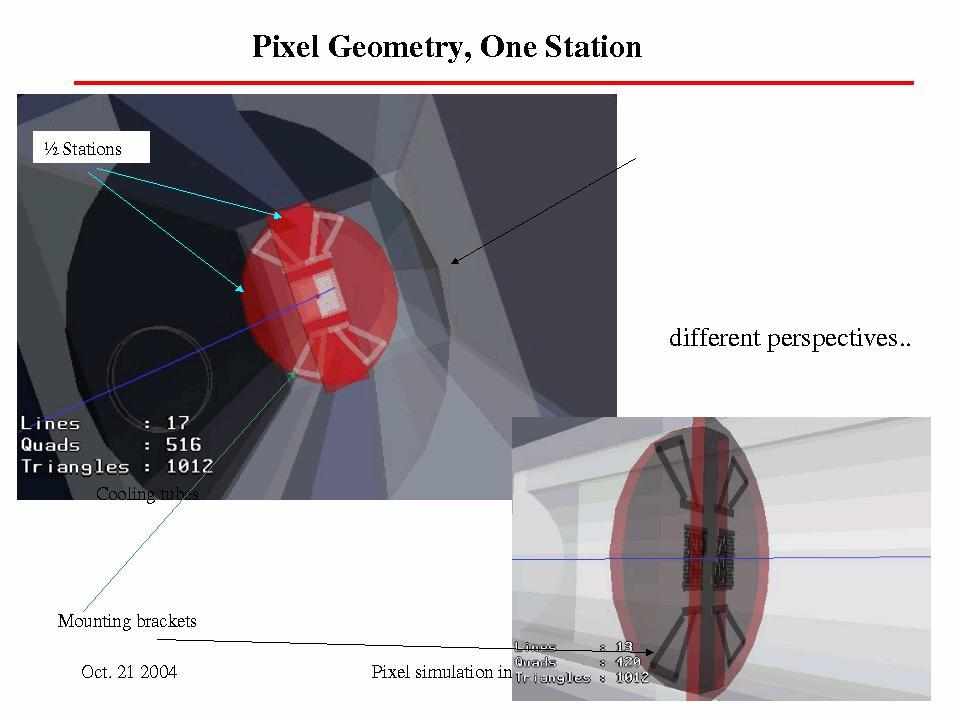}
\caption{\small
Geant4 using the VRML driver is used to produce a picture \newline
 of the proposed BTeV pixel detector.
 }
\label{figure:paul_6}
\end{figure}

       Here is a slide from the BTeV collaboration (Paul Lebrun)
       showing a pixel detector
       Fig.~\ref{figure:paul_6}. This Geant4 slide is produced using the VRML
       driver.

\begin{figure}[htbp]
\includegraphics[width=0.8\textwidth, clip]{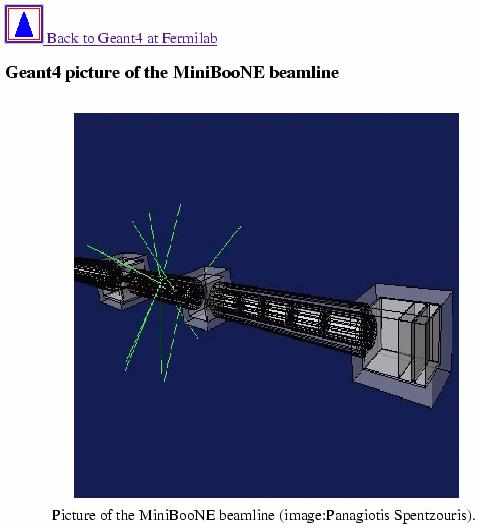}
\caption{\small
Geant4 using the VRML driver is used to produce a picture of the MiniBooNE beamline.
 }
\label{figure:MiniBooNE_beamline}
\end{figure}

        The BTeV experiment was canceled in Feb. 2005\cite{BTeV}.
       An experiment which is currently running (2005) is MiniBooNE.
       This is an experiment to test for neutrino mass by searching
       for neutrino oscillations\cite{Neutrino_mass}.
       Their GEANT4 simulation uses the VRML driver.
       A simulation of their beamline is shown in
       Fig.~\ref{figure:MiniBooNE_beamline}.

\subsection{OpenInventor}
      Although I have decided not to include the OpenInventor driver
       in our discussions,
      I was very impressed with the following figure
      (Fig. \ref{figure:OpenInventor_1})
      \begin{figure}[htbp]
\includegraphics[width=0.8\textwidth, clip]{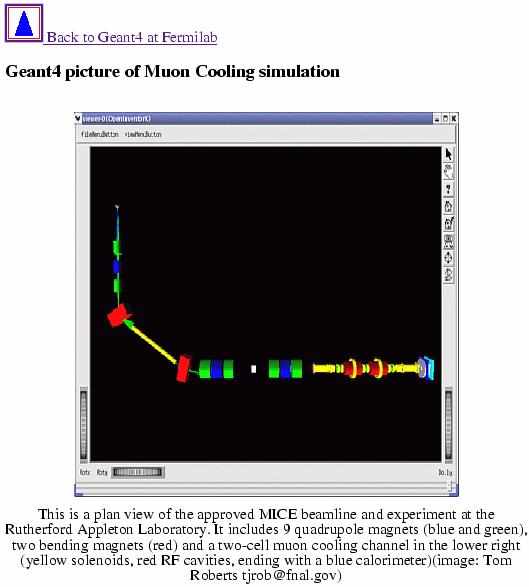}
\caption{\small
Simulation of MICE experiment using OpenInventor as the device driver (Tom
Roberts).
 }
\label{figure:OpenInventor_1}
\end{figure}

      provided by Tom Roberts. This figure is also available from the Fermilab
      Geant4 web page. 
      
\section{Building OpenGL}  
      In order to use some Visualization Drivers an external library
      must be used. There are five such libraries (OpenGL-Xlib,
      OpenGL-Motif, OpenInventor-X, Dawn-Network, and VRML-NetWork).
      Only one of then is among the 4 drivers we are investigating in this
      note (OpenGL-Xlib). The OpenGL version will be built for Fermilab users
      starting with v4\_7\_1. The build was done with the following
      command:\newline
      setenv G4VIS\_BUILD\_OPENGLX\_DRIVER 1\newline
      To setup this version:\newline
      setup geant4 v4\_7\_1  -q GCC\_3\_4\_3-OpenGL \newline
      You should get the correct flavor for your machine
      (either Scientific Linux, or Red Hat Linux).
      To correctly use this version you will need
      to indicate that you wish to use OpenGL and to obtain the libraries.\newline
      setenv G4VIS\_USE\_OPENGLX 1 
 
 	setenv OGLHOME /usr/X11R6\newline
	setenv OGLFLAGS "-I\$OGLHOME/include"\newline
	setenv OGLLIBS  "-L/usr/X11R6/lib -lGL -lGLU"\newline
	
\section{Producing a plot from OpenGL}
The first step is to produce an executable. In our example we use
novice example N03. The instructions on doing this are documented
on our web page\cite{web_instructions}.

\begin{figure}[htbp]
\includegraphics[width=0.8\textwidth, clip]{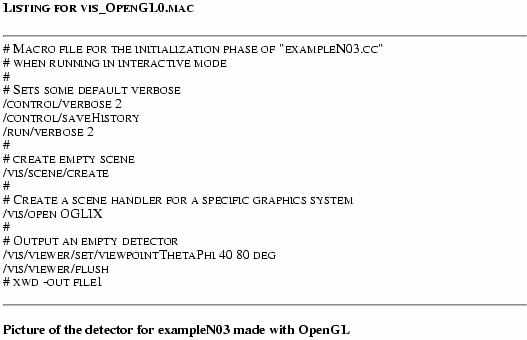}
\caption{\small
Example of a macro file used to produce an OpenGL plot.
The plot shows the detector of novice example N03.
 }
\label{figure:OpenGL_macro_0}
\end{figure}

Next we  produce a picture of the detector. Note this is the sampling
calorimeter with layers of Pb and liquid Ar detection gaps.
We run the job in interactive mode (./exampleN03).
To clearly indicate the commands I am using I have put them in a macro file.
We next execute the macro\newline
Idle$>$ /control/execute vis\_OpenGL0.mac \newline
The macro file is shown in  Fig. \ref{figure:OpenGL_macro_0}.

\begin{figure}[htbp]
\includegraphics[width=0.8\textwidth, clip]{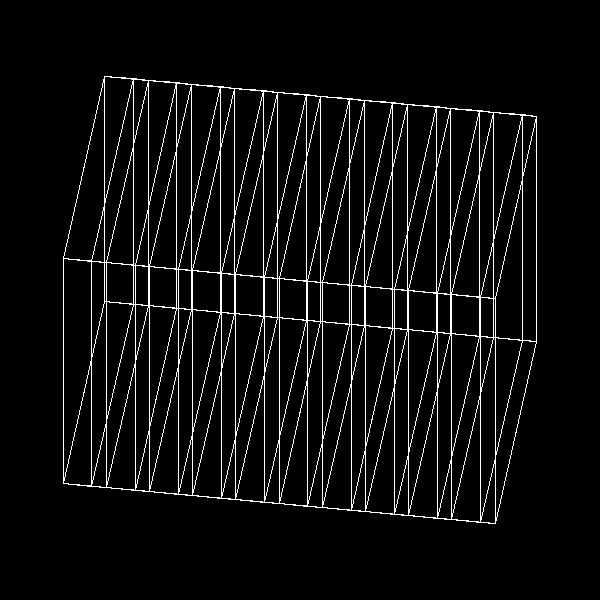}
\caption{\small
The calorimeter of example N03 produced using OpenGL.
 }
\label{figure:ex3_detector_OpenGL0}
\end{figure}

This produces a new window with the picture of the detector.
A saved image of the picture is produced by opening another window.
The  window which has just been opened should be also in the area 
which contains the executable.
In this window you type\newline
 xwd -out file.xwd\newline
The name of the file is arbitrary. I will refer to this as an xwd file. 
This will produce a cursor, which
you should place over the window which contains the picture.
The program convert is then used to change the format of the file.\newline
convert file0.xwd ex3\_detector\_OpenGL0.jpeg\newline
I usually use a jpeg file for the web and an eps file for latex documents. 
The  resulting plot is shown in Fig. \ref{figure:ex3_detector_OpenGL0}. 

\begin{figure}[htbp]
\includegraphics[width=0.8\textwidth, clip]{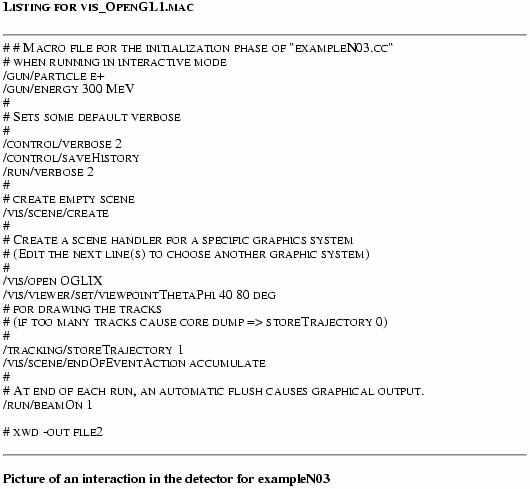}
\caption{\small
How to produce a plot using OpenGL for an interaction in the calorimeter.
 }
\label{figure:OpenGL_macro_1}
\end{figure}

The size of file0.xwd is 704 KBytes,
 and the size of the corresponding jpeg file is 51 KBytes.
 The size of the xwd file is determined by the number of pixels in the window.
We now make a few changes in the macro and draw a plot for an interaction
in the calorimeter. The few changes include specifying the incident particle
and it's energy. The macro file is given in Fig.~\ref{figure:OpenGL_macro_1}.

\begin{figure}[htbp]
\includegraphics[width=0.8\textwidth, clip]{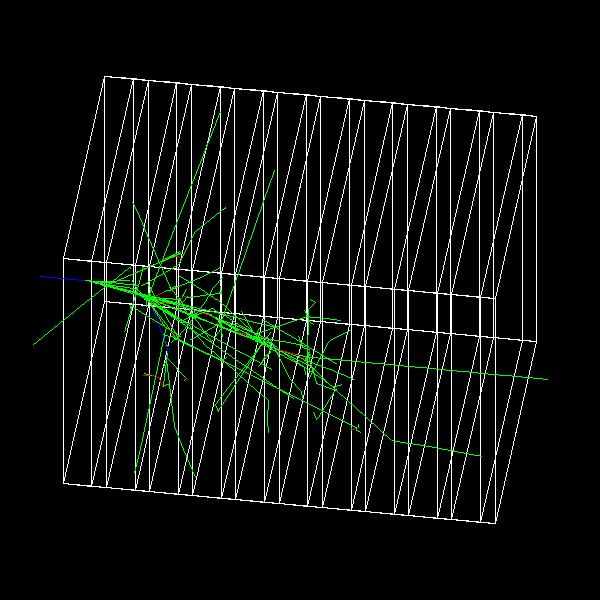}
\caption{A positron of 300 MeV incident on the detector of example N03 (OpenGL).}
\label{figure:ex3_detector_OpenGL1}
\end{figure}

The resulting plot is shown in Fig.~\ref{figure:ex3_detector_OpenGL1}.

\begin{figure}[htbp]
\includegraphics[width=0.8\textwidth, clip]{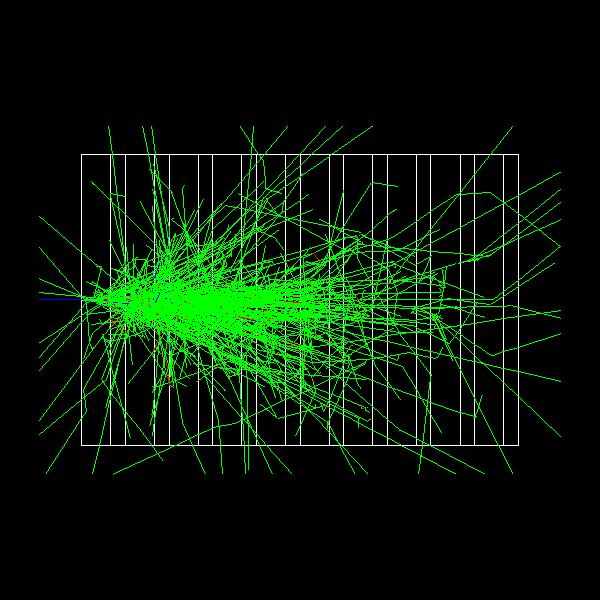}
\caption{\small
The superposition of 10 events in the detector of example N03.\newline
Each event is initiated by a 300 MeV positron (OpenGL). 
 }
\label{figure:ex3_detector_OpenGL2}
\end{figure}

On the screen the figure looks much better, because the green contrasts much
better with the black (the black is not as intense as in the hardcopy).
Geant4 uses the convention that negatively charged tracks are plotted in red,
neutral in green and positive in blue.
The size of the file is again the same (704 KBytes) as expected and the size of the
corresponding jpeg file is 64 KBytes. We now run a third job
accumulating all the hits for 10 events. The size of the file is again the same
(704 KBytes) and the size of the jpeg file is 70 KBytes.
The resulting very complicated plot
is shown in Fig.~\ref{figure:ex3_detector_OpenGL2}.

The size of the corresponding eps files is in all three cases 2144 KBytes.
The information on the sizes of the OpenGL files is summarized in
Table:~\ref{table:size_OpenGL}.
\begin{table}[thbp]
\begin{center}
\caption{Sizes in KBytes for the OpenGL files}
\label{table:size_OpenGL}  
\begin{tabular}{cccc} \\ \hline\hline  
%\multicolumn{4}{c} {Data Sets} \\
   \multicolumn{1}{c} {Name}    &   \multicolumn{1}{c} {xwd}  &
   \multicolumn{1}{c} {jpeg}    &   \multicolumn{1}{c} {eps }       \\[0.05in]
  Detector                 &    704             &    51   &  2144    \\[0.05in]                                                
  Detector + 1 event       &    704             &    64   &  2144    \\[0.05in]
  Detector + 10 events     &    704             &    70   &  2144   \\[0.05in]
      \hline\hline
\end{tabular}
\end{center}
\end{table}

\section{Producing many plots from OpenGL}
The procedure described in the last section works well for a few plots,
but real projects  involve many plots. I am told by Don Holmgren
that the Cryogenic Dark Matter people have developed a procedure
using OpenGL and a virtual window to do this automatically.	
	
\section{Using intermediate files}
In principle you can use all drivers simultaneously. 
I have not yet tried this! First you must indicate which
driver you are using. For the three drivers we will examine in detail
the command is:\newline
setenv G4VIS\_USE\_HepRepFile 1 \newline
setenv G4VIS\_USE\_DAWNFILE 1\newline
setenv G4VIS\_USE\_VRMLFILE 1 \newline

The intermediate files have unique names in that the file is given
an extension (.heprep, prim, wrl) for (HepRepFile, DAWNFILE, VRMLFILE)
respectively.

\subsection{Using HepRep}
I run a single macro file that is almost equivalent to the three macro
file used for OpenGL. The macro as expected uses /vis/open HepRepFile.
This produces three data files
(G4Data0.heprep, G4Data1.heprep and G4Data2.heprep). 
The size of these data files are 71, 1705, 16626 KBytes.

\begin{figure}[htbp]
\includegraphics[width=0.8\textwidth, clip]{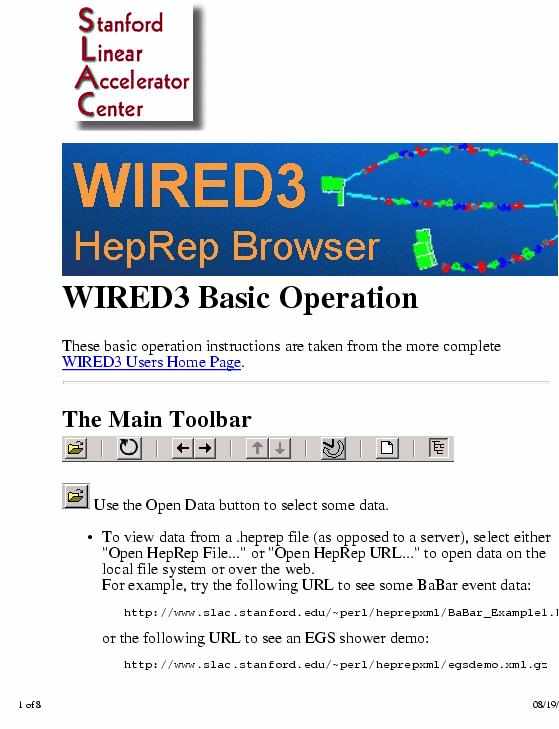}
\caption{\small
The first page of help for the HepRepBrowser..
 }
\label{figure:HepRep_Browser}
\end{figure}

 In order to proceed to look at these
files the HepRep client file WIRED must be installed.
The instruction for installing WIRED are given on a SLAC web page\cite{SLAC_WIRED}.
I have installed wired in my own area on the FNALU
Linux cluster and also on the cepa cluster (~andy/Wired/bin/wired).
Additional information that maybe useful can be found in the Fermilab
dictionary of Geant4 terms\cite{Dictionary_WIRED}.
You will also want to learn about the HepRep Browser.
To bring up the browser type:\newline
~andy/Wired/bin/wired data\_file\cite{data_file}\newline
Information about the browser is available under Help -$>$ Quick Browser
the first page is shown in Fig.~\ref{figure:HepRep_Browser}.

\begin{figure}[htbp]
\includegraphics[width=0.8\textwidth, clip]{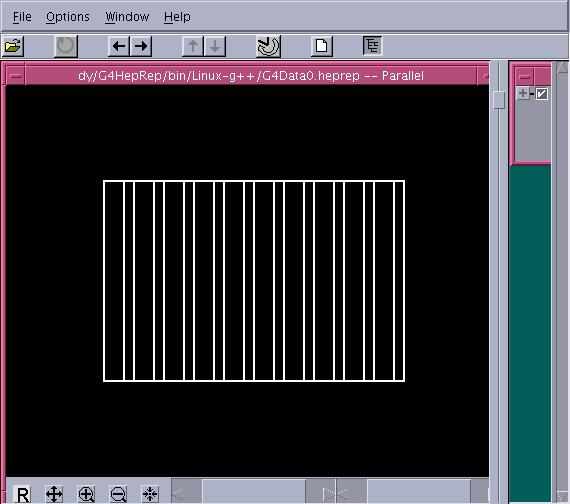}
\caption{\small
The calorimeter of example N03 produced using HepRep.
 }
\label{figure:ex3_HepRep0}
\end{figure}

\begin{figure}[htbp]
\includegraphics[width=0.8\textwidth, clip]{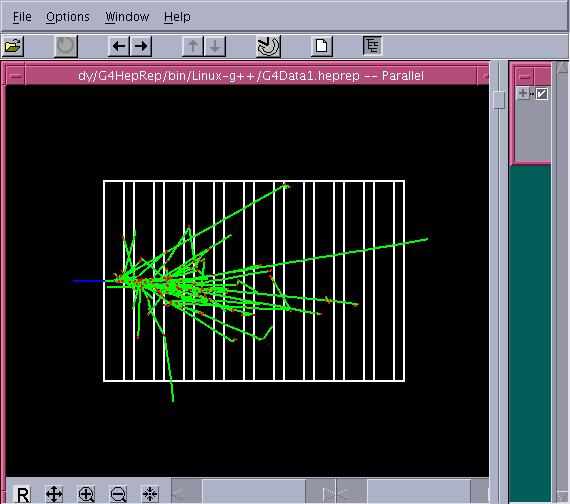}
\caption{\small
A positron of 300 MeV incident on the detector of example N03 (HepRep).
 }
\label{figure:ex3_HepRep1}
\end{figure}

\begin{figure}[htbp]
\includegraphics[width=0.8\textwidth, clip]{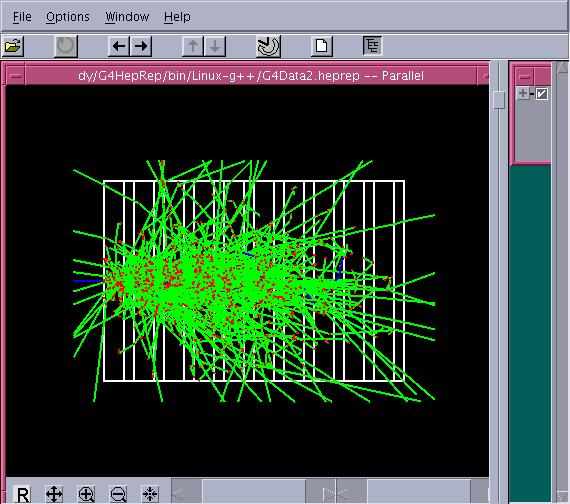}
\caption{\small
The superposition of 10 events in the detector of exampleN03.\newline
Each event is initiated by a 300 MeV positron (HepRep).} 
\label{figure:ex3_HepRep2}
\end{figure}

The plots showing the detector, an interaction in the calorimeter
and the superposition of 10 interactions are shown in Figures
~\ref{figure:ex3_HepRep0},~\ref{figure:ex3_HepRep1} and 
~\ref{figure:ex3_HepRep2}.
 On immediately notices that the 
command /vis/viewer/set/viewpointThetaPhi is not implemented, but one sees
a standard view of the detector. I should point out that rotations
around the axis perpendicular to the screen are allowed
and more (see information in the Quick Browser under "The Orientation
Toolbar"). 

The  way to produce
a plot is to go to "File" and then "Export Graphics". I have recently
encountered some difficulties with this and thus I have produced
plots with xwd (see section on  Producing a plot from OpenGL).
HepRep keeps track of the files so that you can just click on
the right arrow on the Main tool bar and it proceeds from
the first file to the second file (G4Data0.heprep -$>$ G4Data1.heprep).

It takes a relatively long time
(82 sec on my machine which is a 500 MHZ Pentium III\cite{MHZ}) to go
from the second plot to the third This is clearly caused by the 
large size of the file and the fact that HepRep requires much information to
be stored in it's menus. 
As indicated earlier other information
can be learned from the plots. For example that there are 469
tracks in the event see Fig.~\ref{figure:ex3_HepRep1}. 
 
\begin{figure}[htbp]
\includegraphics[width=0.8\textwidth, clip]{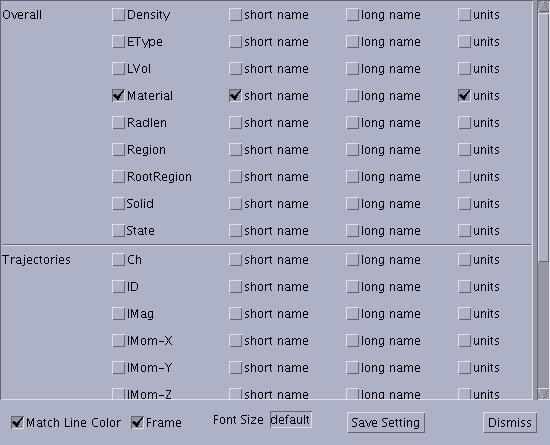}
\caption{\small
HepRep allows many options to be set including material.
 }
\label{figure:ex3_HepRep1_label}
\end{figure}

To set the material 
go to "Options" and then "Label Control" (see
Fig~\ref{figure:ex3_HepRep1_label}). 

\begin{figure}[htbp]
\includegraphics[width=0.8\textwidth, clip]{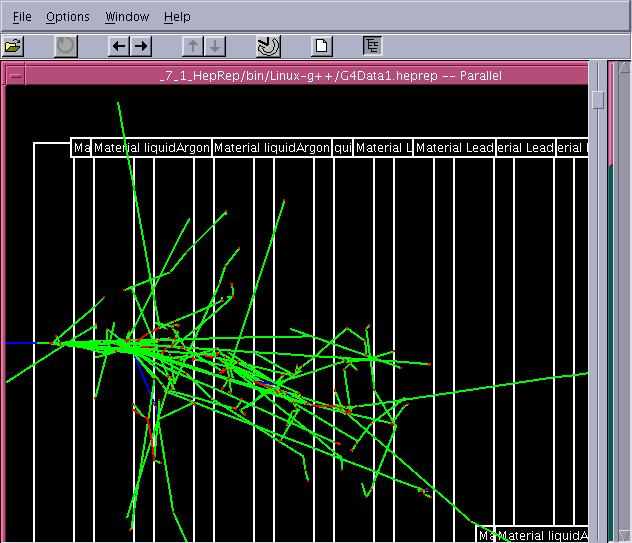}
\caption{\small
HepRep plot with the short version of the label for material.
 }
\label{figure:ex3_HepRep1_material}
\end{figure}

The resulting plot is shown
in Fig.~\ref{figure:ex3_HepRep1_material}. 

One can see that the materials
"lead: and "liquid Argon" appear on the plot, but to which region they refer is
not so clear.

\subsection{Using DAWN}
The same macro is run as for HepRep, except for one command
/vis/open DAWNFILE. This produces three data files (g4\_00.prim, g4\_01.prim
and g4\_02.prim). The size of the data files are 12, 235, and
2267 KBytes. Considerable smaller than the sizes of the HepRep files.

Again to proceed further the DAWN client must be installed.
This is available form the web site of Satoshi Tanaka one of it's developers\cite{Tanaka}.
To install you will need to answer some questions, I think
the default answer will work. 
On the fnalu cluster(flxi02,...) it is installed in ~andy/dawn/dawn
and on the cepa cluster in ~andy/G4TEST/dawn\_3\_85e/PRIM\_DATA/dawn.

\begin{figure}[htbp]
\includegraphics[width=0.8\textwidth, clip]{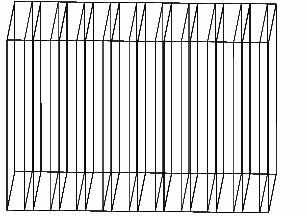}
\caption{\small
The calorimeter of example N03 produced using DAWN.
 }
\label{figure:ex3_DAWN0}
\end{figure}

The plots showing the detector, an interaction in the calorimeter
and the superposition of 10 interactions are shown in Figures
\ref{figure:ex3_DAWN0},

\begin{figure}[htbp]
\includegraphics[width=0.8\textwidth, clip]{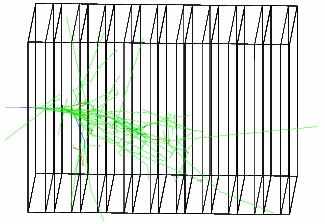}
\caption{\small
A positron of 300 MeV incident on the detector of example N03 (DAWN).
 }
\label{figure:ex3_DAWN1}
\end{figure}

\begin{figure}[htbp]
\includegraphics[width=0.8\textwidth, clip]{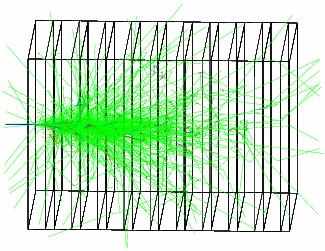}
\caption{\small
The superposition of 10 events in the detector of exampleN03.\newline
Each event is initiated by a 300 MeV positron (DAWN). 
}
\label{figure:ex3_DAWN2}
\end{figure}

~\ref{figure:ex3_DAWN1}

 and 
~\ref{figure:ex3_DAWN2}. 

One immediately notices that the 
command /vis/viewer/set/viewpointThetaPhi is now implemented.

\begin{figure}[htbp]
\includegraphics[width=0.8\textwidth, clip]{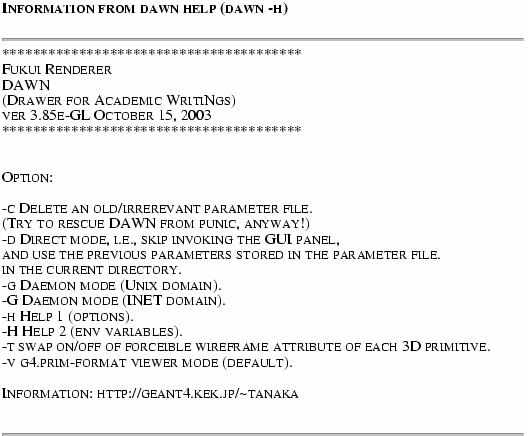}
\caption{\small
The help file for the DAWN display.
 }
\label{figure:DAWN_help}
\end{figure}

The DAWN job runs immediately because of the smaller sized files
and that no menu information is needed.
What immediately means is 5 sec for the 10 event file. In running
DAWN I have used the -d option this is for the direct mode. Further information
is available dawn -h (see Fig.~\ref{figure:DAWN_help}).

\subsection{Using VRML}
The same macro is run as for HepRep, except for one command
/vis/open VRML1FILE. This produces three data files (g4\_00.wrl, g4\_01.wrl
and g4\_02.wrl). The size of the data files are 6, 141, and
1370 KBytes. Considerable smaller than the sizes of the HepRep files.

Again to proceed further the VRML client must be installed. Information
about VRML can be found on the web\cite{3D_consortium}. It is produced by
"Web3D Consortium- Open Standards for Real-Time 3D Communication
Creators of the VRML (VRML1, VRML97, and now X3D, previously called
VRML200x) open standards". On cepa cluster the VRML client
is in\newline  
/home/prj/bphys/releases/dev/vrmlview. 
On the fnalu cluster(flxi02,...) it is installed in ~andy/vrmlview/vrmlview.	
	
\begin{figure}[htbp]
\includegraphics[width=0.8\textwidth, clip]{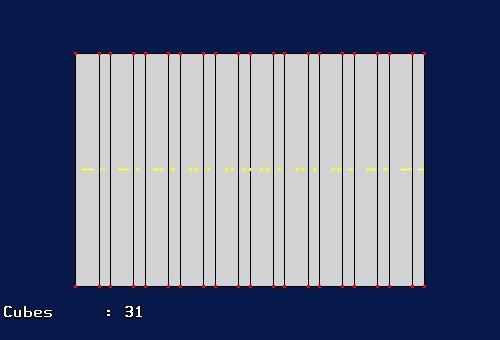}
\caption{\small
The calorimeter of example N03 produced using VRML.
 }
\label{figure:ex3_VRML0}
\end{figure}

The plot showing the detector is in Fig.~\ref{figure:ex3_VRML0}
and was produced by going to File and then  "Save snapshot".

\begin{figure}[htbp]
\includegraphics[width=0.8\textwidth, clip]{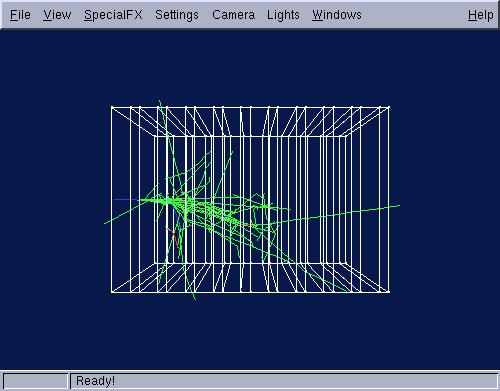}
\caption{\small
A positron of 300 MeV incident on the detector of example N03 (VRML).
 }
\label{figure:ex3_VRML1b}
\end{figure}

We continue with our standard Fig.~\ref{figure:ex3_VRML1b} 
of a positron entering the calorimeter.

\begin{figure}[htbp]
\includegraphics[width=0.8\textwidth, clip]{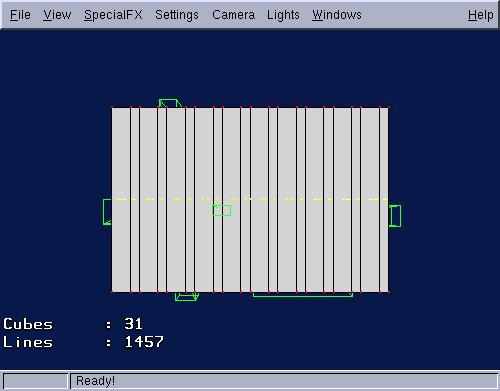}
\caption{\small
A positron of 300 MeV incident on the detector of example N03 (VRML).
One can easily set the defaults such that the display is not useful.
 }
\label{figure:ex3_VRML1a}
\end{figure}

 This plot was produced by selecting under "View"
"Wireframe" and "Vertices". However, it is easy to select options that
do not produce a useful figure (see Fig. ~\ref{figure:ex3_VRML1a}).

\begin{figure}[htbp]
\includegraphics[width=0.8\textwidth, clip]{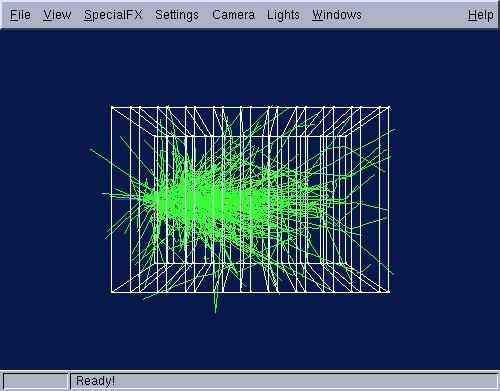}
\caption{\small
The superposition of 10 events in the detector of exampleN03.\newline
Each event is initiated by a 300 MeV positron (VRML). 
}
\label{figure:ex3_VRML2a}
\end{figure}

Our standard final plot is shown in Fig.~\ref{figure:ex3_VRML2a}. 

It takes
50 seconds for this plot to appear.

\subsection{Summary of intermediate files}
The size of the intermediate data files is summarized 
in Table ~\ref{table:size}.
\begin{table}[thbp]
\begin{center}
\caption{Sizes in KBytes for the intermediate files}
\label{table:size}  
\begin{tabular}{cccc} \\ \hline\hline  
%\multicolumn{4}{c} {Data Sets} \\
   \multicolumn{1}{c} {Name}    &   \multicolumn{1}{c} {HepRep}  &
   \multicolumn{1}{c} {DAWN}    &   \multicolumn{1}{c} {VRML }       \\[0.05in]
  Detector                 &     71             &    12   &     6    \\[0.05in]                                                
  Detector + 1 event       &   1705             &   235   &   141    \\[0.05in]
  Detector + 10 events     &  16626             &  2267   &  1370   \\[0.05in]
      \hline\hline
\end{tabular}
\end{center}
\end{table}

We see that if size is your major factor then VRML is the best choice and
HepRep is the worst. In terms of menus to assist you HepRep is the best, 
the next best is VRML and last is DAWN. In terms of time DAWN is a clear winner
5 sec versus 50 sec for VRML and 82 sec for HepRep.In terms of quality plots
DAWN is again the clear winner, followed by VRML and then HepRep.

\section{Which driver to use?}
\begin{figure}[htbp]
\includegraphics[width=0.8\textwidth, clip]{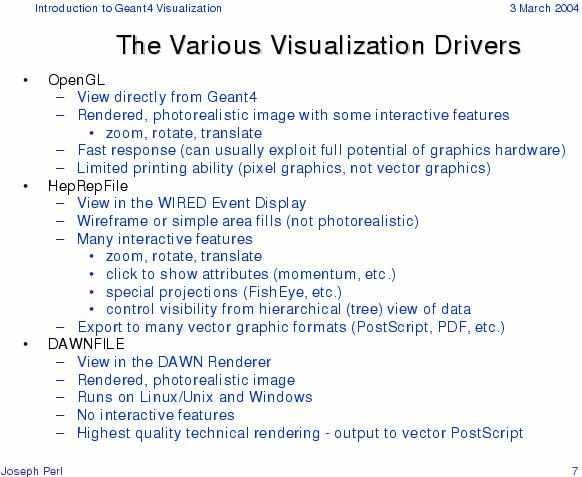}
\caption{\small
Features of the Visualization Drivers, OpenGL, HepRep and DAWN
 }
\label{figure:G4V7}
\end{figure}

\begin{figure}[htbp]
\includegraphics[width=0.8\textwidth, clip]{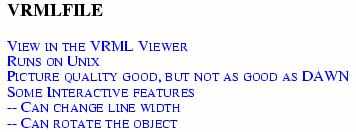}
\caption{\small
Features of the VRML Visualization Driver
 }
\label{figure:VRML_info}
\end{figure}

The question of which driver to use was addressed by Joseph Perl.
First he answers the list of what features are associated
with each driver(see Fig~\ref{figure:G4V7}).
 To complete his slide I present the
results for VRML in Fig.~\ref{figure:VRML_info}.

\begin{figure}[htbp]
\includegraphics[width=0.8\textwidth, clip]{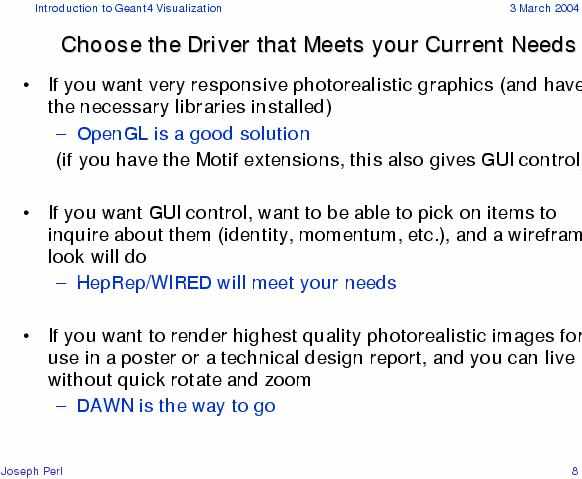}
\caption{\small
When to use a given driver from  Joseph Perl SLAC Tutorial.
 }
\label{figure:G4V8}
\end{figure}

 Perl also indicates 
the circumstances  in which you want 
to use the different drivers  in Fig.~\ref{figure:G4V8}.

At Fermilab we have three different types of users, large experiments (CMS)
medium experiments (MiniBooNE) and small experiments (neutron therapy).

The large experiments will want a structured environment. They may
however, build it out of an environment that already has many of the tools
like HepRep or they may want to proceed from a less structured environment
like VRML. I strongly suspect that they will want their own graphics
program. The Geant4 convention of only three colors (red, green, blue)
for (neg, neutral, pos) is not what is needed for a modern high
energy simulation program.
One clearly needs different colors for electron and muons etc.
The advice for the builders of a display
for the Geant4 simulation program is to proceed exactly the way Geant4 has done
and produce a program that is device independent! It would be even better 
if the builders of such a display program could get thir code incorporated
into Geant4.
 
I suspect that medium sized experiments will want to go with HepRep
because of the assistance provided by the menu's.

Small experiments and starting students will want to use OpenGL
because of it's quick response. It also my favorite in the spirit of keeping
things simple.
 
Clearly DAWN may be needed for publication quality work by large, medium and
small experiments.

In Table\ref{table:summary_all}
\begin{table}[thbp]
\begin{center}
\caption{Graphics Summary 
}
\label{table:summary_all}  
\begin{tabular}{cccccc} \\ \hline\hline  
\multicolumn{6}{c} {(1)= Best, (2) = Next Best} \\
   \multicolumn{1}{c} {}        &   \multicolumn{1}{c} {1}    &
   \multicolumn{1}{c} {2}       &   \multicolumn{1}{c} {3}    &
   \multicolumn{1}{c} {4}       &   \multicolumn{1}{c} {5}         \\[0.05in]
   Quality of plots    &    DAWN      &   OpenGL   &  VRML       &  HepRep  &        \\[0.05in]                                                
   Menus               &   HepRep     &   VRML     &  DAWN       &  OpenGL  &        \\[0.05in]
   Time                &   OpenGL     &   DAWN     &  VRML       &  HepRep  &        \\[0.05in]
   Size of Files       & OpenGL(jpeg) &   VRML     & OpenGL(eps) &  DAWN    & HepRep \\[0.05in]
      \hline\hline
\end{tabular}
\end{center}
\end{table}

 I have attempted to summarized
some of the important features for determining which device driver is
right for you. The comparison of time and size of files is somewhat
arbitrary when one compares the direct device driver and the those
that use an intermediate file. I have left out the time required
to produce the initial file for (HepRep, DAWN, VRML). Similarly
I have not included the time to run OpenGL and produce the xwd file.

All four drivers are available on the central cluster, and the Geant4-team
will try to assist you in obtaining the different drivers\cite{geant4-users}.
The issue of support
has changed dramatically recently because of a suggestion by GP Yeh.
It is now possible to run Geant4 on your own personal computer (Linux pc)\cite{geant4_pc}.
For pc user they will need to know how to obtain the client software.
As this is all free software no serious problems are expected.

\section{Acknowledgments}

	I thank Randy Herber for his help with Latex.

% \input{t_data_sets}
%  \input{ch_2}
  
% \input{t_Monte_Carlo}
%  \input{ch_3} 

%4 
% \input{t_backgrounds}
 % \input{ch_4}
% \input{t_new_physics}

% \input{t_conclusion}
%  \input{ch10n}
  
%\input{ack_vis}
%\section{Acknowledgements}

%We would like to thank XXX for helpful comments.
%\section{Acknowledgments}

%	I thank Randy Herber for his help with Latex.

 %\input{t_app_A}
%\input{ref_vis}
%\input{t_tables}
%\input{fig_vis}


\begin{thebibliography} {99}
\bibitem{OSCAR}
OSCAR\_3\_9\_6 is based on Geant v4\_6\_2\_p02. \newline
Unfortunately it still uses some of the old Geant3 code:\newline
libG3Main.so and libG3Interface.so
\bibitem{RayTracer} 
At Fermilab, setup Geant4 version \newline
 setup	geant4 v4\_7\_1 -q GCC\_3\_4\_3 and then go to \newline
/products/geant4source/v4\_7\_1/NULL/geant4.7.1/source/visualization/RayTracer/README.
\bibitem{data_file}
The data files are produced in the same directory as the exe file.\newline
It is convenient to set a variable to point to this area.\newline
setenv G4EXE ~andy/geant4/G4TEST\_4\_7\_1\_HepRep\newline
data\_file -$>$ \$G4EXE/G4Data0.heprep
\bibitem{geant4-users}
People who want to be informed about Geant 4 activities
at Fermilab should subscribe to the FNAL mailing list GEANT4-USERS.
\end{thebibliography}
\end{document}